\documentclass[conference,compsoc]{IEEEtran}
\ifCLASSOPTIONcompsoc
  \usepackage[nocompress]{cite}
\else
  \usepackage{cite}
\fi
\usepackage{graphicx}
\usepackage[caption=false, font=footnotesize]{subfig}
\usepackage{tabularx}
\usepackage{booktabs}
\newcolumntype{Y}{>{\centering\arraybackslash}X}

\ifCLASSINFOpdf
\else
\usepackage[caption=false, font=footnotesize]{subfig}

\fi
\usepackage{algorithm}
\usepackage{algorithmic}

\usepackage{lipsum}

\begin{document}
\renewcommand{\algorithmicrequire}{\textbf{Input:}}
\renewcommand{\algorithmicensure}{\textbf{Output:}}
\title{Wave Function Collapse Coloring: A New Heuristic for Fast Vertex Coloring}

\author{\IEEEauthorblockN{Anh Mac}
\IEEEauthorblockA{American School of Warsaw \\
Warsaw, Poland\\
Email: 22mac\_a@aswarsaw.org}
\and
\IEEEauthorblockN{David Perkins}
\IEEEauthorblockA{Hamilton College\\
New York, USA\\
Email: dperkins@hamilton.edu}}


%


\maketitle

\begin{abstract}
In this paper, we propose a high-speed greedy sequential algorithm for the vertex coloring problem (VCP), based on the Wave Function Collapse algorithm, called Wave Function Collapse Coloring (WFC-C). An iteration of this algorithm goes through three distinct phases: vertex selection, color restriction through wave function collapsing, and domain propagation. In effect, WFC-C propagates color choices or "domain" restrictions beyond immediate neighbourhoods. This heuristic, combined with a series of other greedy optimizations, allows for a fast algorithm that prevents certain color conflicts. Through extensive experiments, we show that WFC-C remains competitive (and occasionally better) in terms of optimal coloring, and dramatically outperforms existing high-speed VCP, with on average speed differences ranging from 2000 times to 16000 times, on the most difficult instances of the DIMACS benchmark. 
\end{abstract}


%
\IEEEpeerreviewmaketitle

\section{Introduction}
The Vertex Coloring Problem (VCP), a sub-problem of the Graph Coloring Problem, is an NP-hard combinatorics optimization problem with a wide range of applications, studied extensively in literature. VCP asks, in essence, to assign a color to every vertex such that no adjacent vertex shares a color. A common extension of the VCP is to find the minimum number of colors to create a valid coloring, called the chromatic number $\chi(G)$. Examples of this problem's applications include frequency assignment in networks \cite{park1996application, Castelino1996}; timetabling \cite{sabar2012graph, miner1995optimizing}; register allocation in compilers \cite{chaitin1982register, briggs1994improvements}. See \cite{formanowicz2012survey} or \cite{ahmed2012applications} for a survey on the applications of VCP. While exact approaches to solving the VCP exist \cite{sewell1996improved, mendez2006branch, eppstein2002small, zhou2014exact}, they are impractical for real-life applications as exact algorithms are unable to solve large graphs due to the amount of time required. Thus, researchers tend to concentrate on heuristic solutions. Traditionally, heuristic and metaheuristic algorithms for VCP can be split into three distinct categories: constructive approaches \cite{brelaz1979new, leighton1979graph, Culberson95exploringthe}; local searching (including simulated annealing \cite{kose2017simulated}, quantum annealing \cite{titiloye2011quantum}, tabu search \cite{hertz1987using, dorne1999tabu, galinier2006survey}, variable neighborhood searching \cite{avanthay2003variable}); and population-based approaches \cite{porumbel2010evolutionary, fleurent1996genetic, hindi2012genetic, chalupa2011population, galinier1999hybrid}. More recently, modern approaches have incorporated machine and statistical learning techniques. For example, \cite{goudet2020populationbased} introduces a population-based approach with gradient descent optimization, and \cite{zhou2018improving} uses probability learning on a local search algorithm to produce more optimal colorings. An exhaustive study of popular heuristic methods can be found in \cite{sun2018heuristic}. 
\par However, with the exception of sequential construction algorithms, modern literature places an emphasis on optimal coloring as opposed to time efficiency. Despite this focus on optimal coloring, fast graph coloring is essential in a large number of applications, such as computing the upper bounds in branch-and-bound algorithms for the maximum cliche problem \cite{san2011exact, batsyn2014improvements}, or to use graph coloring-based compression techniques to speed up automatic differentiation \cite{gebremedhin2005color, hovland2005sensitivity}. Many hybrid VCP algorithms use fast but inaccurate vertex coloring algorithms to generate a high estimate of the chromatic number and repeatedly lower this until a legal $k$-coloring cannot be reached while other algorithms optimize the initial, inaccurate coloring directly \cite{zhou2018improving, titiloye2011quantum, porumbel2010evolutionary, hertz1987using, dorne1999tabu, avanthay2003variable}. In such applications, speed is more important than achieving an optimal coloring. Despite modern literature's focus on optimal VCP algorithms, high-speed vertex coloring is still vital to many crucial applications. 
\par Approaches to high-speed VCP solutions generally consist of greedy and constructive algorithms. These algorithms iterate through a set of all vertices, assigning a color following some rules until a valid coloring is reached. Once a coloring is assigned, it is not reconsidered. Most effective high-speed VCP algorithms employ a dynamic ordering of vertices to produce more optimal coloring. Famous examples of these high-speed VCP algorithms are \textit{maximum saturation degree} (DSatur) \cite{brelaz1979new}, \textit{recursive largest first} (RLF) \cite{leighton1979graph}, and the iterated greedy algorithm (IG) \cite{Culberson95exploringthe}. These are the algorithms we compare our novel algorithm to. More recently, an algorithm proposed by \cite{komosko2016fast} implements a greedy-style algorithm using bit-wise operations to increase time efficiency. However, the majority of these fast VCP solutions do not restrict colors of vertices beyond the immediate neighborhood, nor is there any metaheuristic processing to optimize coloring.  
\par In this paper, we present a fast heuristic vertex coloring algorithm, hereafter called Wave Function Collapse Coloring (WFC-C). The key contribution provided in WFC-C is the propagation of color restrictions beyond the immediate neighborhood. Computational results show that WFC-C dramatically outperforms existing fast VCP algorithms in speed, with an average speed increase ranging from 2,662 to 16,124 times faster, while producing optimally competitive colorings.
\par The structure of the paper is as follows: in Section 2, we discuss the formal definition of the Graph Coloring Problem; in Section 3, we outline previous fast VCP algorithms and their variations; in Section 4, we present the formal algorithm for WFC-C; in Section 5, we generate experimental results and compare such results to existing literature; in Section 6, we conclude with a discussion on the uses of WFC-C and possible future research regarding WFC-C.

\section{Problem Description}
Let $G = (V, E)$ be a simple, undirected graph with vertex set $V = \{1, 2, ..., n\}$ and edge set $E \subset V \times V$. Given $G$, find a $k$ and a mapping $c : V \to \{1, ..., k \}$ such that $c(i) \neq c(j)$ for each edge $(i, j) \in E$. The optimization version of VCP asks to find the minimum $k$, denoted as the chromatic number $\chi(G)$. 

\begin{figure}[!h]
\centering
\includegraphics[width=0.5\columnwidth]{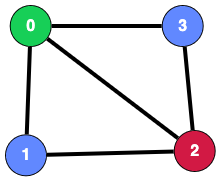}
\caption{A proper vertex coloring}
\label{fig_valid}
\end{figure}
\par The classical problem asks for the minimum amount of colors needed such that no adjacent vertices share the same color. Figure \ref{fig_valid} is an example of a proper coloring. The classical problem is connected to the $k$-coloring problem ($k$-GCP), which asks if a graph can be colored in $k$ colors. The classic problem is proven to be NP-hard, whereas the $k$-coloring problem is NP-complete \cite{gareyjohnson1990}.  

\section{Fast Heuristic Approaches} 
\label{sec-fast}
In this section, we discuss three algorithms: \textit{iterated greedy}, \textit{maximum saturation degree}, and \textit{recursive largest first}. Section \ref{sec-experiments} compares the performance of our algorithm to the performances of these.

\subsection{Iterated Greedy}
The \textit{Iterated Greedy} algorithm (IG) proposed by \cite{Culberson95exploringthe} consists of assigning each vertex in a static ordering its first available color. As such, a valid coloring can be found in, on average, linear time and in the worst cases, in exponential time, resulting in one of the fastest VCP algorithms. Algorithm 1 provides pseudocode for the IG algorithm. 
 \begin{algorithm}[H]
 \caption{Iterated Greedy (IG)}
 \begin{algorithmic}[1]
    \REQUIRE{A simple connected graph $G = (V, E)$, with all vertices uncolored, and list of distinct colors}
    \ENSURE{A proper vertex coloring}
    \STATE $\omega \leftarrow$ highest-degree first ordering of $V$
    \FOR{vertex $u$ in $\omega$}
        \STATE assign to $u$ the smallest color that is not already used by one of $u$'s neighbors
    \ENDFOR
 \end{algorithmic}
 \end{algorithm}
 
 Naturally, there always exists some ordering which provides the optimal coloring \cite{husfeldt2015graph}. Research tends to focus on discovering the optimal ordering \cite{husfeldt2015graph, kuvcera1991greedy}. While there is no general consensus of which ordering is optimal, most modern implementations of IG uses a \textit{highest-degree first} ordering, as high-degree vertices are often the ones with the most potential for color conflict.
\subsection{Maximum Saturation Degree}
\textit{Maximum Saturation Degree} (DSatur), as written in \cite{brelaz1979new}, is a dynamic greedy coloring algorithm. We define \textit{saturation} as the number of colored vertices in a vertex's neighborhood. The first vertex colored is one with the highest degree. As vertices are colored, DSatur selects an uncolored vertex with the highest saturation, and colors it with the first available color. Ties between vertices with the same saturation are broken by choosing the vertex with the higher degree. Algorithm 2 provides psuedocode for DSatur. 
\begin{algorithm}[H]
 \caption{Maximum Saturation Degree (DSatur)}
 \begin{algorithmic}[1]
    \REQUIRE{A simple connected graph $G = (V, E)$, with all vertices uncolored}
    \ENSURE{A proper vertex coloring}
    \STATE $\omega \leftarrow$ highest-degree first ordering of $V$
    \WHILE{$\omega$ is not empty}
    \STATE $u \leftarrow$ a vertex with maximum saturation, breaking ties using a vertex of highest degree among these
    \STATE assign to $u$ the smallest color that is not already used by one of $u$'s neighbors 
    \STATE remove $u$ from $\omega$
    \ENDWHILE
 \end{algorithmic}
 \end{algorithm}
\subsection{Recursive Largest First} 
\textit{Recursive Largest First} (RLF), first introduced by \cite{leighton1979graph}, constructs a coloring by creating stable color sets of each color. RLF starts by assign a color $k$ to the uncolored vertex with the highest degree. Then, all neighbors of that vertex are moved to a separate set. The uncolored vertex that is adjacent to the most vertices in this separate set is assigned color $k$, and all of its neighbors are moved to this separate set. This process is repeated until all uncolored vertices in this separate set are assigned a color, at which point the entire algorithm is repeated with a new color (thus, the algorithm is recursive). Algorithm 3 provides an iterative version of RLF.
\begin{algorithm}
 \caption{Recursive Largest First (RLF) (Iterative)}
 \begin{algorithmic}[1]
    \REQUIRE{A simple connected graph $G = (V, E)$, with all vertices uncolored}
    \ENSURE{A proper vertex coloring}
    \STATE $W \leftarrow V$
    \STATE $k \leftarrow 0$
    \WHILE{$W$ is not empty}
        \STATE $U \leftarrow W$
        \STATE $W \leftarrow \emptyset$
        \STATE $v \leftarrow$ the vertex with the highest degree in $U$, breaking ties at random
        \STATE assign color $k$ to $v$
        \STATE remove $v$ from $U$
        \STATE move all neighbors of $v$ from $U$ into $W$ 
        \WHILE{$U$ is not empty}
            \STATE $u \leftarrow$ the vertex in $U$ with the most neighbors in $W$, breaking ties at random
            \STATE assign color $k$ to $u$
            \STATE remove $u$ from $U$
            \STATE move all neighbors of $u$ from $U$ into $W$ 
        \ENDWHILE
        \STATE $k \leftarrow k + 1$
    \ENDWHILE
 \end{algorithmic}
 \end{algorithm}
  Computational studies show that RLF consistently produces more optimal coloring than DSatur, however, RLF is consistently slower \cite{chiarandini2010analysis}. 
\section{WFC-C: A New Coloring Heuristic} 
Our new vertex coloring heuristic, Wave Function Collapse Coloring (WFC-C), is a greedy non-backtracking coloring algorithm with a minimum remaining values heuristic based on the Wave Function Collapse (WFC) algorithm proposed by \cite{Gumin_Wave_Function_Collapse_2016}. WFC is a procedural image generation algorithm used to generate bitmaps and tiles from small source images. Given a source image, the WFC algorithm extracts a set of patterns that it then uses to paint a canvas, starting with a random selection of pattern and location. From that starting point, the algorithm \textit{observes} the canvas in order to find the location that can accept the fewest patterns, \textit{collapses} that location by choosing one of those patterns at random, and \textit{propagates} the effects of that choice throughout the local neighborhood. The names of the functions in our pseudocode reflects these steps.
\par It may be helpful to compare our algorithm to its counterpart in quantum physics, thinking of each uncolored vertex as existing in a superposition of all available colors. By collapsing the wave function of any particular vertex, the colors available to neighboring vertices are further restricted to adhere to the requirements of the vertex coloring problem. 
\par We define the \textit{domain} of a vertex as a list of colors still available to it, and the \textit{entropy} of the vertex as the size of its domain. At each iteration, our algorithm selects the vertex with the lowest entropy, greedily assigns that vertex the lowest available color, and restricts the domains of neighbors (and their neighbors, and so on, if necessary). 
\par A reasonable initial number of available colors derives from Brooks' theorem \cite{Lovsz1975}, which states that for almost all simple, connected graphs $G$, the upper bound for the chromatic number $\chi(G)$ is $\triangle(G)$ (the degree of the vertex that has maximum degree). We therefore initialize our algorithm with a list of $\triangle(G)$ colors.

\par Our pseudocode begins in Algorithm 4. In it, we let $M = \triangle(G)$ be the number of colors available when the algorithm first runs; if the algorithm encounters a state where a vertex cannot be legally colored, it increments $M$ and restarts. Let $L = [1, 2, \ldots, M]$ be a list of distinct colors. We consider the variables $M$ and $L$ to be global.
\par We define $\texttt{domain}(u)$ to be those colors still available to the vertex $u$ as the algorithm proceeds, and $\texttt{entropy}(u) = | \texttt{domain}(u) |$. Finally, we use $\texttt{color}(u)$ to denote the color assigned by the algorithm to vertex $u$. 
\begin{algorithm}
 \caption{WFC-C: Wave Function Collapse Coloring}
 \begin{algorithmic}[1]
    \REQUIRE{A simple connected graph $G = (V, E)$, with all vertices uncolored}
    \ENSURE{A proper vertex coloring}
    \STATE $\texttt{color}(u) \leftarrow 1$ for any vertex $u$ such that $deg(u) = M$
    \STATE propagate($u$)
  \WHILE{there are uncolored vertices}
  \STATE $v \leftarrow$ observe($G$)
  \STATE collapse($v$)
  \STATE propagate($G, v$)
  \ENDWHILE
  \STATE output the valid coloring
 \end{algorithmic}
 \end{algorithm}
 To get started, we assign a color to the vertex with highest degree, as any delay in coloring this vertex will likely increase the chance that we will need to use an unused color when we reach it.
 \begin{algorithm}[H]
 \caption{observe($G$)}
 \begin{algorithmic}[1]
    \REQUIRE{A graph $G = (V, E)$}
    \ENSURE{The vertex $v$ with lowest non-zero entropy}

    \STATE $v \leftarrow$ the uncolored vertex with lowest entropy
    \IF{$\texttt{entropy}(v) = 0$}
        \STATE $M \leftarrow M + 1$
        \STATE restart the algorithm
    \ENDIF
    \STATE return $v$
 \end{algorithmic}
 \end{algorithm}
The function \textit{observe()} returns the vertex that has lowest entropy. If any vertex has entropy $0$, there are no available colors for that vertex, and we have reached an illegal coloring. When this happens, the algorithm restarts. Once all vertices are properly colored, the algorithm outputs the valid coloring.
\begin{algorithm}[H]
 \caption{collapse($v$)}
 \begin{algorithmic}[1]
    \REQUIRE{The vertex $v$ selected by \texttt{observe()}}
    \STATE $\texttt{color}(v) \leftarrow \min( \texttt{domain}(v) )$
 \end{algorithmic}
 \end{algorithm}

\par The function \textit{collapse()} greedily assigns the lowest available color to a vertex. The analog equivalent can be found, again, in physics, whereby observing a particle, the wave function of that particle collapses. While in physics, the propagation of restriction of a particle's domain impacts entangled particles instantaneously, our algorithm calculates the impact of the domain restriction in the function \textit{propagate()}. 
 \begin{algorithm}[H]
 \caption{propagate($G, v$)}
 \begin{algorithmic}[1]
    \REQUIRE{A graph $G = (V, E)$ and vertex $v \in V$}
    \STATE create an empty stack $A$
    \STATE $\texttt{push}(A, v)$
    \WHILE{$A$ is not empty} 
        \STATE $u \leftarrow \texttt{pop}(A)$
        \FOR{each uncolored neighbor $v$ of $u$ with $\texttt{entropy}(v) > 1$}
            \STATE remove $\texttt{color}(u)$ from $\texttt{domain}(v)$ if possible
            \IF{$\texttt{entropy}(v) = 1$ and $v \not \in A$} 
                \STATE $\texttt{color}(v) \leftarrow$ the only color in $\texttt{domain}(v)$
                \STATE $\texttt{push}(A, v)$
            \ENDIF
        \ENDFOR
    \ENDWHILE
 \end{algorithmic}
 \end{algorithm}

 The function \textit{propagate()} determines the impact of assigning a color to a vertex (in the \textit{collapse()} function) on the vertices in the neighborhood of that vertex. When the entropy of a vertex falls to $1$, we add it to the stack $A$, triggering a further domain restriction on the neighbors of that vertex. This cascading effect lasts until we no longer have vertices with entropy $1$.
\begin{figure}[!h]
\label{fig_propagate}
\centering
\subfloat[Unaffected domain]{%
       \includegraphics[width=0.45\linewidth]{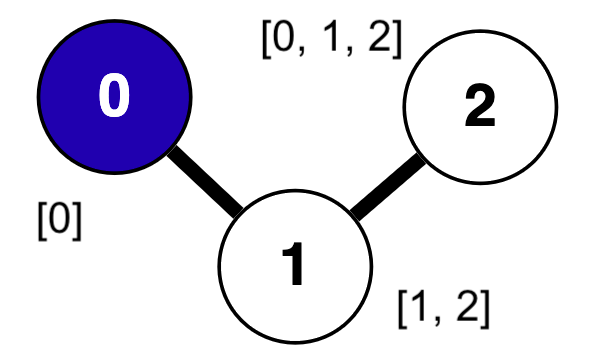}
    }
    \hfill
\subfloat[Propagated domain]{%
       \includegraphics[width=0.45\linewidth]{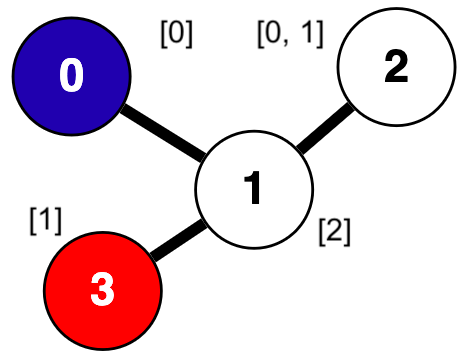}
    }
    \hfill
\caption{Two examples of domain propagation}
\end{figure}
\par Consider the graph on the left in Figure \ref{fig_propagate} that has vertices numbered $0, 1, 2$. Vertex $0$ has just been colored blue. The lists beside each vertex are their domains (with $0$ as blue, $1$ as red, and $2$ as some other as-yet unused color). The domain of vertex $1$ is restricted by the color of vertex $0$, but this restriction is not passed to vertex $2$ (because the entropy of vertex $1$ has not fallen to $1$). Contrast this situation with the graph on the right in Figure \ref{fig_propagate}, where vertex $1$ has a red neighbor that, along with its blue neighbor, has entropy $!$. The \textit{propagate()} function therefore restricts the domain of vertex $2$ by removing color $2$.

\par This produces a vertex coloring algorithm that, in worst cases, runs in exponential time. Nevertheless, this heuristic approach yields a dramatic increase in speed in comparison to other high-speed VCP algorithms. Propagating domain restrictions beyond immediate neighborhoods leads to a more efficient colorings, as fewer color conflicts are created. 

\par
\begin{figure}[!h]
\centering
\subfloat[Coloring by a greedy algorithm]{%
       \includegraphics[width=0.45\linewidth]{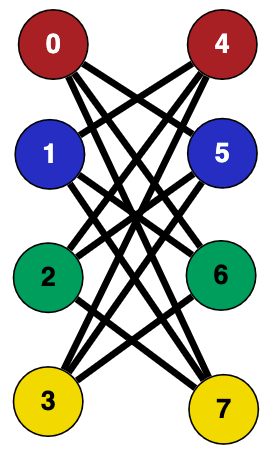}
    }
    \hfill
\subfloat[Coloring by WFC-C]{%
       \includegraphics[width=0.45\linewidth]{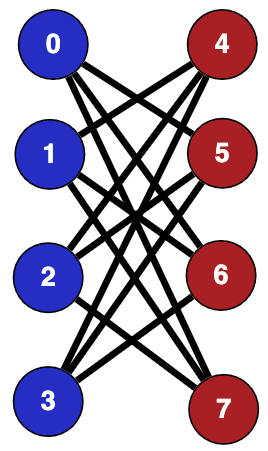}
    }
    \hfill
\caption{Colorings of an 8 vertex crown graph}
\label{fig_crown}
\end{figure}

 Choosing the next vertex to color based on entropy rather than on a predetermined ordering of the vertices allows WFC-C to optimally color graphs that are known to produce non-optimal outputs with greedy methods. For example, it is well known that crown graphs (also known as Johnson's graphs) lead to worst-case outputs in greedy VCP algorithms \cite{johnson1974worst}. A crown graph (see Figure \ref{fig_crown}) is a complete bipartite graph with vertex sets $\{u_{1}, u_{2}, \ldots, u_{n}\}$ and $\{v_{1}, v_{2}, \ldots, v_{n}\}$ with every edge ($u_{i}$, $v_{i}$) removed. Figure 3 shows the colorings found by a greedy algorithm and by WFC-C. 

\begin{table*}[!t]
\renewcommand{\arraystretch}{1.3}
\caption{Comparison between WFC-C and other well known high-speed VCP algorithms}
\label{table_example}
\centering
\begin{tabularx}{\textwidth}{lYYYYYYYc}
\toprule
& \multicolumn{2}{c}{WFC-C} & \multicolumn{2}{c}{IG \cite{Culberson95exploringthe}} & \multicolumn{2}{c}{DSatur \cite{brelaz1979new}} & \multicolumn{2}{c}{RLF \cite{leighton1979graph}} \\
\cmidrule(lr){2-3}
\cmidrule(lr){4-5}
\cmidrule(lr){6-7}
\cmidrule(lr){8-9}

Graph ($k^{*}) $ & $k$ & Time ($\mu$s) & $k$ & Time ($\mu$s) & $k$ & Time ($\mu$s) & $k$ & Time ($\mu$s)\\
\cmidrule{1-9}
dsjc250.5 (28) & 37 & 13.512 & 43 & 11115.007 & 41 & 23751.985 & 36 & 42914.899\\
\cmidrule(lr){2-3}
\cmidrule(lr){4-5}
\cmidrule(lr){6-7}
\cmidrule(lr){8-9}
dsjc500.1 (12) & 16 & 29.879 & 21 & 33605.821 & 19 & 69989.930 & 15 & 388956.352 \\
\cmidrule(lr){2-3}
\cmidrule(lr){4-5}
\cmidrule(lr){6-7}
\cmidrule(lr){8-9}
dsjc500.5 (48) & 65 & 29.853 & 75 & 46869.319 & 72 & 111675.122 & 61 & 305158.286\\
\cmidrule(lr){2-3}
\cmidrule(lr){4-5}
\cmidrule(lr){6-7}
\cmidrule(lr){8-9}
dsjc500.9 (126) & 163 & 77.567 & 182 & 58027.525 & 173 & 142268.025 & 160 & 146150.221\\
\cmidrule(lr){2-3}
\cmidrule(lr){4-5}
\cmidrule(lr){6-7}
\cmidrule(lr){8-9}
dsjc1000.1 (20) & 26 & 68.645 & 33 & 141975.182 & 30 & 341228.525 & 24 & 2718414.290\\
\cmidrule(lr){2-3}
\cmidrule(lr){4-5}
\cmidrule(lr){6-7}
\cmidrule(lr){8-9}
dsjc1000.5 (83) & 117 & 73.018 & 131 & 194245.460 & 124 & 506310.930 & 111 & 2295954.474\\
\cmidrule(lr){2-3}
\cmidrule(lr){4-5}
\cmidrule(lr){6-7}
\cmidrule(lr){8-9}
dsjc1000.9 (222) & 307 & 61.501 & 331 & 236598.067 & 318 & 649709.885 & 291 & 1208017.936\\
\cmidrule(lr){2-3}
\cmidrule(lr){4-5}
\cmidrule(lr){6-7}
\cmidrule(lr){8-9}
le450\_15c (15) & 24 & 28.664 & 35 & 28962.309 & 27 & 61993.171 & 23 & 340777.315\\
\cmidrule(lr){2-3}
\cmidrule(lr){4-5}
\cmidrule(lr){6-7}
\cmidrule(lr){8-9}
le450\_15d (15) & 25 & 26.962 & 36 & 28853.385 & 27 & 57664.130 & 23 & 254909.280\\
\cmidrule(lr){2-3}
\cmidrule(lr){4-5}
\cmidrule(lr){6-7}
\cmidrule(lr){8-9}
le450\_25c (25) & 29 & 27.245 & 42 & 29649.291 & 31 & 64280.814 & 28 & 204600.458\\
\cmidrule(lr){2-3}
\cmidrule(lr){4-5}
\cmidrule(lr){6-7}
\cmidrule(lr){8-9}
le450\_25d (25) & 29 & 26.121 & 43 & 29964.561 & 31 & 57866.094 & 29 & 249406.028 \\
\cmidrule(lr){2-3}
\cmidrule(lr){4-5}
\cmidrule(lr){6-7}
\cmidrule(lr){8-9}
flat300\_28\_0 (28) & 42 & 16.649 & 48 & 20113.007 & 46 & 34017.846 & 38 & 126560.001\\
\cmidrule(lr){2-3}
\cmidrule(lr){4-5}
\cmidrule(lr){6-7}
\cmidrule(lr){8-9}
flat1000\_76\_0 (82) & 115 & 60.869 & 126 & 191775.106 & 125 & 494438.279 & 108 & 2271508.909\\
\cmidrule(lr){2-3}
\cmidrule(lr){4-5}
\cmidrule(lr){6-7}
\cmidrule(lr){8-9}
r1000.5 (234) & 247 & 61.114 & 400 & 191942.573 & 250 & 477752.565 & 253 & 2824829.388\\
\cmidrule(lr){2-3}
\cmidrule(lr){4-5}
\cmidrule(lr){6-7}
\cmidrule(lr){8-9}
dsjr500.5 (122) & 127 & 29.325 & 199 & 52389.390 & 133 & 111122.395 & 135 & 331169.317\\
\cmidrule(lr){2-3}
\cmidrule(lr){4-5}
\cmidrule(lr){6-7}
\cmidrule(lr){8-9}
dsjr500.1c (85) & 92 & 30.456 & 106 & 61360.008 & 103 & 161692.195 & 97 & 114366.285\\
\cmidrule(lr){2-3}
\cmidrule(lr){4-5}
\cmidrule(lr){6-7}
\cmidrule(lr){8-9}
r250.5 (65) & 68 & 13.444 & 105 & 10938.443 & 71 & 23176.511 & 68 & 40086.790\\
\cmidrule(lr){2-3}
\cmidrule(lr){4-5}
\cmidrule(lr){6-7}
\cmidrule(lr){8-9}
r1000.1c (98) & 111 & 64.263 & 134 & 245390.317 & 122 & 652403.469 & 119 & 606413.984\\
\cmidrule(lr){2-3}
\cmidrule(lr){4-5}
\cmidrule(lr){6-7}
\cmidrule(lr){8-9}
latin\_square (98) & 132 & 56.007 & 213 & 177341.010 & 213 & 411207.398 & 131 & 1700047.135\\
\cmidrule(lr){2-3}
\cmidrule(lr){4-5}
\cmidrule(lr){6-7}
\cmidrule(lr){8-9}
C2000.5 (148) & 209 & 124.225 & 230 & 833020.021 & 221 & 2083126.550 & N/A & N/A\\
\cmidrule(lr){2-3}
\cmidrule(lr){4-5}
\cmidrule(lr){6-7}
\cmidrule(lr){8-9}
C4000.5 (272) & 379 & 249.987 & 409 & 3233834.482 & 400 & 8532030.081 & N/A & N/A\\
\bottomrule

\end{tabularx}
\end{table*}

\section{Experimental Results}
\label{sec-experiments}

Here we compare WFC-C's performance to the algorithms described earlier (in Section \ref{sec-fast}): iterated greedy (IG), maximum saturation degree (DSatur), and recursive largest first (RLF). 

\subsection{Experimental Environment}
We implemented this algorithm in Python 3.7, on a 1.8 GHz Dual-Core Intel i5 processor. To test the performance of WFC-C, we used a popular graph coloring benchmark from the second DIMACS Implementation Competition \cite{DIMACS1996}. DIMACS has previously been used as the primary comparison point between a wide variety of vertex coloring algorithms. \cite{kose2017simulated, titiloye2011quantum, porumbel2010evolutionary, hindi2012genetic, chalupa2011population, galinier1999hybrid, goudet2020populationbased, zhou2018improving, hutter2014algorithm}. Graphs from DIMACS include: Leighton graphs \textit{le}; random graphs \textit{dsjc}; flat graphs \textit{flat}; geometric random graphs labelled \textit{dsjr} and \textit{r} (these graphs can also have suffix \lq \lq c", meaning the compliment of their respective graphs); large random graphs $C2000.5$ and $C4000.5$; and a Latin square graph. There is also a standard benchmark for comparing performances on different hardware.\footnote{https://mat.gsia.cmu.edu/COLOR03/} Our machine reported a user time of 12.07 seconds on the r500.5.b test. While the DIMACS benchmark has many more graphs than is shown in Table 1, we focus on the \lq \lq difficult" instances, as is customary in modern literature \cite{titiloye2011quantum, kose2017simulated, porumbel2010evolutionary}. 
\par In Table 1, we use the term $k^{*}$ to represent the best upper bound on the number of colors that has currently been proposed in literature. These $k^{*}$ values were achieved using optimal, but slower algorithms, on graphs whose chromatic number $\chi(G)$ is unknown or has not been proven.    
\par We compared WFC-C to other popular fast heuristic solutions, including IG, DSatur and RLF. For IG, we used a \textit{highest-degree first} ordering. We implemented each algorithm, ran each algorithm 100 times, and took the average time. The exact code for replication of this experiment has been made public.\footnote{https://github.com/LightenedLimited/WaveFunctionCollapse-Coloring}

\subsection{Results}
An important result to note is the \lq \lq N/A" for RLF on the graphs $C2000.5$ and $C4000.5$. These were graphs where the RLF algorithm did not complete the experiment on our machine within an hour time window  (an estimation based on one iteration of $C4000.5$ suggests that RLF would have taken over three hours to finish the experiment).
\par Evidently, WFC-C is dramatically faster than any of the compared algorithms. In comparison to the \textit{Iterated Greedy} algorithm, our algorithm colored the same graphs 2,662 times faster on average. In the case of $C4000.5$, our algorithm was 12,936 times faster than IG. As for DSatur, our algorithm performed even faster in comparison. WFC-C averaged 6,582 times faster, and in the case of $C4000.5$, WFC-C was 34,130 times faster. Of the graphs in which RLF completed, WFC-C performed on average 16,124 times faster. In the case of $r1000.5$, WFC-C performed 46,222 times faster than RLF.  
\par In terms of the coloring produced, WFC-C produces more optimal coloring in all instances than IG and DSatur. Over the course of the entire dataset, WFC-C produced 21.04\% more optimal coloring than IG. In the case of $r1000.5$, WFC-C produced a coloring that was 38.25\% more optimal. As for DSatur, WFC-C is 9.22\% more optimal and in the case of $latin\_square$, WFC-C was 38.03\% more optimal. As mentioned in Section 4, WFC-C produces more optimal results likely due to the propagation of domain restrictions beyond the immediate neighborhood. This propagation heavily impacts the color assignments and prevents color conflicts by restricting color domains a conflict occurs.   
As for RLF, WFC-C produces equally or more optimal coloring on larger graphs, whereas RLF produces more optimal coloring on smaller graphs. Of the 6 instances where WFC-C produced an equal or more optimal coloring ($le450\_25d$, $r1000.5$, $dsjr5005.5$, $dsjr500.1c$, $r250.5$, $r1000.1c$), 5 occurred in larger graphs. As for the differences in coloring for small, they're relatively minor, with the majority of instances being a difference of one or two $k$-values. A possible explanation behind this behavior is the viewing manner in restricting the domain of vertices. WFC-C uses a stack to store vertices whose domain restricts its neighbors, resulting in a depth-first propagation of domain restriction. In larger graphs, this depth-first propagation may result in more efficient color assignments and prevent conflicts from occurring. However, this depth-first behavior may not be as efficient as RLF's restiction pattern, which is akin to breadth-first viewing, for smaller graphs. Since smaller graphs will inevitably cause color conflicts faster, WFC-C's depth-first propagation may cause conflicts faster or, at the very least, prevent less conflicts than RLF's breadth-first style of domain restriction. 
\par In general, none of the presented algorithms reach the best-known upper-bound on the chromatic number ($k^{*}$) in any instance. That is due to the fact that algorithms used to reach these upper-bounds are not fast VCP algorithms, but rather specialized, time-consuming optimization solutions. For example, the $k^{*}$ of $dsjc1000.9$ was reached by a quantum annealing solution proposed in \cite{titiloye2011quantum}. This solution took 13,740 seconds on a 3 GHz Intel processor, whereas WFC-C took $6.1501 \cdot 10^{-6}$ seconds on a 1.8 GHz Intel processor. It is unrealistic to expect sequential, high-speed VCP solutions to approach these best-known upper bounds. 
\section{Conclusion and Further Research}
In this paper, we presented a high-speed VCP algorithm which greedily propagates domain restrictions of vertices beyond immediate neighbourhoods. This propagation strategy allows our Wave Function Collapse Coloring (WFC-C) algorithm, loosely based on \cite{Gumin_Wave_Function_Collapse_2016}, to successfully prevent color conflicts and allows for a drastic increase in speed in comparison to other popular fast VCP solutions. 
Extensive computational studies with difficult instances of the DIMACS benchmark show that WFC-C significantly outperforms existing GCP approaches in speed, with average speed increases ranging from 2,000 times to 16,000 times, and competes competitively with existing sequential algorithms in terms of coloring efficiency.
\par As for future work, this domain restriction approach is likely applicable in many other combinatorial optimization problems of similar nature.  Furthermore, we only considered this algorithm's value as a solution to the VCP. The main purpose of the original WFC is the generation of an image given a source bitmap \cite{Gumin_Wave_Function_Collapse_2016}. With minor modifications, our algorithm can perform a similar function (that is to say, coloring a larger graph in a similar fashion as a source graph). By changing the domain requirements and propagation requirements, we can assign colors to vertices in successive and restricting patterns and in appropriate frequencies found in a source graph. This, essentially allows us to color a larger graph given a source graph, similar to how WFC generated a larger image given a source image.

\bibliographystyle{IEEEtran}  
\bibliography{bibi}  


\end{document}